\documentclass[useAMS,usenatbib]{mn2e}

\usepackage{amssymb}
\usepackage{amsmath}
\usepackage{times}

\newcommand{\lya}{Ly$\alpha$}
\newcommand{\Lya}{\mbox{Ly$\alpha$}}
\newcommand{\Ha}{\mbox{H$\alpha$}}
\newcommand{\HI}{H\,{\sc i}}
\newcommand{\nhi}{\mbox{$N_{\rm HI}$}}
\newcommand{\msun}{\mbox{$M_\odot$}}
\newcommand{\mhi}{\mbox{$M_{\rm HI}$}}
\newcommand{\kms}{\mbox{km s$^{-1}$}}
\newcommand{\cm}{cm$^{-2}$}

\newcommand{\degree}{\hbox{$^\circ$}}
\newcommand{\h}{$h^{-1}_{100}$}
\newcommand{\civ}{\hbox{C\,{\sc iv}}}
\newcommand{\mgii}{\hbox{Mg\,{\sc ii}}}
\newcommand{\ovi}{\hbox{O\,{\sc vi}}}

\title[Absorber-galaxy cross-correlation]
{Cross correlation of Lyman-$\alpha$ absorbers with gas-rich galaxies}

\author[E.V. Ryan-Weber]{Emma V. Ryan-Weber$^1$\thanks{email:
	eryan@ast.cam.ac.uk}\\ 
	$^1$Institute of Astronomy, University of Cambridge, Madingley Rd,
	Cambridge CB3 0HA, UK}

\begin{document}

\date{Accepted 2006 January 3. Received 2005 December 20; in original
form 2005 October 17}

\pagerange{\pageref{firstpage}--\pageref{lastpage}} \pubyear{2006}

\maketitle

\label{firstpage}

\begin{abstract}
  The \HI\ Parkes All Sky Survey (HIPASS) galaxy catalogue is
  cross-correlated with known low redshift, low column density (\nhi\
  $<10^{15}$ \cm) Lyman-$\alpha$ absorbers from the literature. The
  redshift-space correlation is found to be similar in strength to
  HIPASS galaxy self-clustering (correlation length $s_{0,ag}=6\pm4$
  and $s_{0,gg}=3.1\pm0.5$ \h\ Mpc respectively). In real-space the
  cross-correlation is stronger than the galaxy auto-correlation
  (correlation length $r_{0,ag}=7.2\pm1.4$ and $r_{0,gg}=3.5\pm0.7$
  \h\ Mpc respectively) on scales from $1-10$ \h\ Mpc, ruling out the
  mini-halo model for the confinement \lya\ absorbers at the 99
  percent confidence level. Provided that the cause of the strong
  cross-correlation is purely gravitational, the ratio of correlation
  lengths suggest that absorbers are embedded in dark matter haloes
  with masses log($M$/\msun) $=14.2$ \h, similar to those of galaxy
  groups. The flattening of the cross-correlation at separations less
  than $\sim$ 600 \h\ kpc could correspond to the thickness of
  filaments in which absorbers are embedded. This work provides
  indirect statistical evidence for the notion that galaxy groups and
  large-scale filaments, particularly those that comprise gas-rich
 galaxies, are the dominant environments of low column density \lya\
  absorbers at $z=0$.
\end{abstract}


\section{Introduction}
\label{sec:intro}


At the present epoch less than 10 per cent of baryonic matter is
located in galaxies (Cole et al. \citeyear{Cole01}; Zwaan et
al. \citeyear{Zwaan03}). The majority of baryons therefore lie in the
intergalactic medium (IGM), and knowledge of the distribution of these
baryons is essential to our understanding of how matter is organised
in the Universe. Lyman-$\alpha$ (\lya) absorption towards distant
quasars reveals the presence of neutral hydrogen (\HI) in the IGM. At
low redshift, observations show that these low column density (\nhi\
$<10^{15}$ \cm) absorbers exist in a variety of environments,
including large-scale filaments \citep[e.g.][]{LeBrun96, Penton02,
Rosenberg03}, galaxy groups \citep[e.g.][]{Lanzetta96, Bowen02}, and
even voids \citep{McLin02} and underdense regions \citep{Grogin98}. A
positive connection between absorbers and galaxies has been
established by many studies \citep{Morris93, Lanzetta95, Bowen96,
Tripp98, Chen98, Chen01, Impey99, Ortiz-Gil99}, however most conclude
that a one-to-one correspondence between absorbers and individual
galaxies does not exist. \cite{Cote05} and \cite{Putman06} provide
compelling evidence that absorbers trace the cosmic web, finding that
absorbers at large radii do not follow galactic rotation curves. Thus
the absorbers appear to lie in the same overdensities as the galaxies,
but not in individual galaxy haloes.

Despite the absorbers' positive association with highly biased,
strongly clustered objects, i.e. galaxies and large-scale structures,
low-\nhi\ \Lya\ absorbers are among the weakest self-clustering
objects in the low redshift Universe. The two-point correlation
function in velocity separation, $\Delta v$, shows mild absorber
clustering for $\Delta v\lesssim500$ \kms\ \citep{Tripp98, Impey99,
Dobrzycki02, Penton02}, weaker than that of galaxy self-clustering
\citep[although, see ][]{Ulmer96}. Hydrodynamic simulations also
measure weak clustering of small gas overdensities in velocity space
\citep{Dave03}, these overdensities, $0.5<
\log(\rho_H/\bar{\rho}_H)<1.5$, correspond to \lya\ absorbers with
$13\lesssim\log (\nhi/$\cm$)\lesssim14$ \citep{Dave99}. When compared
with galaxies identified in the simulations,
a filamentary structure emerges, where absorbers of increasing column
density arise in closer proximity to galaxies.

The cross-correlation function of absorbers and galaxies can be used
to measure the extent to which the two populations of objects are
associated. At intermediate redshifts the strong cross-correlation
between Lyman break galaxies and \civ\ absorbers is used to argue that
these two populations are in fact the same object
\citep{Adelberger05}. The spatial co-incidence of \civ\ absorbers and
Lyman break galaxies however can equally be interpreted as
pre-enrichment from dwarf galaxies born from the collapse of
$\sim2\sigma$ fluctuations at $6<z<12$ \citep{Madau01,Porciani05}. The
same degeneracy has the potential to plague the interpretation of the
cross-correlation of absorbers and galaxies at $z=0$. How can we tell
the difference between a strong cross-correlation signal due to
absorbers arising in galaxy haloes and that due to a general
overdensity of matter that surrounds galaxies? This issue is related
to the claim that $10^{14}\lesssim$ \nhi\ $\lesssim10^{18}$ \cm\
absorbers arise within the haloes of galaxies, based on the
anti-correlation between impact parameter and \lya\ absorber
equivalent width \citep{Lanzetta95, Chen98, Chen01}. Solid
observational evidence \citep{Bowen02, Cote05}, numerical simulations
\citep{Dave99} and analytic models \citep{Lin00} have now resolved
this degeneracy, showing that the anti-correlation between
absorber-galaxy separation and density of absorbing material is a
natural consequence of the overdensity of matter that surrounds
galaxies, extending well beyond galaxy haloes. Since we know that
absorbers and galaxies are largely not the same object, their
cross-correlation signal can be used to measure the extent to which
they are associated.  The cross-correlation signal can also, subject
to interpretation, provide an indirect method to measure the bias of
the absorbers, i.e. the relationship between \HI\ column density and
underlying dark matter density.

In the Press--Schechter \citep[PS,][]{Press74} formalism, the ratio of
the bias of two populations of objects is equal to the ratio of the
cross- to auto-correlation function
\citep[e.g.][]{Mo96,Mo02}.\footnote{In this paper the function that
compares the same population of objects is referred to as the
auto-correlation, and the function that compares two populations is
referred to as the cross-correlation.} Therefore, by (i) knowing the
characteristic total (dark plus baryonic) halo mass of a galaxy
population, (ii) measuring the galaxy auto-correlation function, and
(iii) measuring the absorber-galaxy cross-correlation function, the
mass of the haloes in which the absorbers are embedded can be
inferred. This method has been successfully used by
\cite{Bouche04}. They cross-correlated \mgii\ absorbers with Luminous
Red Galaxies (LRGs) to show that the \mgii\ absorbers are embedded in
haloes with masses $\sim2-8\times10^{11}$\msun, consistent with the
expectation that $\sim40$ per cent of \mgii\ absorbers arise in Damped
\lya\ (DLA) systems \citep{Rao06}, and that DLAs are expected to have
total halo masses of that order. DLAs are gravitationally collapsed
objects and at low redshift are consistent with the local galaxy
population weighted by \HI\ cross section \citep{Zwaan05}. Lower column
density absorbers are not collapsed and may not follow the same bias
trends. At intermediate redshifts the distribution of \lya\ absorbers
reveal structures on scales up to 17 \h\ Mpc \citep{Liske00}, at these
redshifts the bias and clustering properties of \lya\ absorbers
are well modelled \citep{Cen98, McDonald02}. However the physics of
the IGM increases in complexity towards $z=0$. As noted by
\cite{Dave03} the correlation function of gas overdensity no longer
follows the correlation function of \HI\ optical depth at $z=0$,
showing that the bias is not as predictable as at higher
redshifts. Some models however place absorbers in individual
mini-haloes, which do have a well defined bias.

Mini-halo models predict that absorbers arise in gravitationally
confined gas within dark matter mini-haloes \citep{Ikeuchi86, Rees86,
Mo94}. Since mini-haloes are less massive than galaxy haloes, the
absorber-galaxy cross-correlation is expected to be weaker than the
galaxy auto-correlation. The mini-halo model is not an ideal
description of \lya\ absorbers. Too many haloes per unit redshift are
required to account for the observed density of absorption lines,
since the mini-haloes have small spatial cross sections. 

The absorber-galaxy cross-correlation function is instrumental in
establishing whether absorbers are associated with individual
mini-haloes, galaxies or large-scale structure. Although other
statistics have been used by other authors, such as nearest neighbour
and galaxy density distributions, the properties of the
cross-correlation function can be related to the underlying dark
matter density. The cross-correlation is also a clean statistic as it
does not rely on assumptions about correlated pairs. \cite{Morris93}
is the only $z=0$ absorber-galaxy cross-correlation function published
to-date that includes a full three-dimensional calculation. They used
17 absorbers along the 3C 273 line-of-sight and found that the
absorber-galaxy cross-correlation is weaker than the galaxy
auto-correlation on scales from $1-10$ $h_{80}^{-1}$ Mpc. \cite{Mo94}
explain this finding with a combination of $20-30$ per cent of
absorbers arising in galaxies and the remainder arising in
mini-haloes.

In this paper the cross-correlation function is calculated using 129
absorbers along 27 lines-of-sight from the literature and 5,317
galaxies from a blind 21-cm emission-line survey. The galaxy and
absorber data are described in Section~\ref{sec:data}. The details of
the cross-correlation and halo-mass calculation methods are described
in Section~\ref{sec:method}. The subsequent results are outlined in
Section~\ref{sec:results}, including the redshift- and real-space
correlation functions. The nature of the \lya\ absorber correlation
signal and its relationship to galaxies, groups of galaxies and
large-scale filaments is discussed in Section~\ref{sec:dis}, with a
final summary given in Section~\ref{sec:summary}. To compare with
correlation functions in the literature $H_0=100$ \kms\ Mpc$^{-1}$ is
used throughout.

\section{Data and Random Samples}
\label{sec:data}
\subsection{HIPASS Galaxies}

The \HI\ Parkes All Sky Survey (HIPASS) is a blind survey for
extragalactic 21-cm emission, thus it provides a census of gas-rich
galaxies at $z=0$. Extragalactic sources in HIPASS have been
catalogued in the southern ($\delta$$<$+2\degree, Meyer et al.,
\citeyear{Meyer04}) and northern (2\degree$<$$\delta$$<$+25.5\degree,
Wong et al. \citeyear{Wong05}) parts of the survey.  The combined
HIPASS catalogues cover 29,579 square degrees and the heliocentric
velocity range from 300 to 12,700 \kms, and contain a total of 5,317
galaxies. HIPASS has a velocity resolution of 18 \kms\ and optical
matching shows a $1\sigma$-positional accuracy of 1.4\arcmin\ (Doyle
et al. \citeyear{Doyle05}), corresponding to less than 4 \h\ kpc at
the minimum absorber distance. The HIPASS galaxy catalogue is
available at {\tt{<www.aus-vo.org>}}.


The uniform coverage of HIPASS over such a large fraction of the sky
makes it an attractive survey for a cross-correlation function, as no
corrections for gaps in the survey are required and edge effects are
minimised. A recent study by \cite{Cooper05} finds that edges and
holes in a survey can have a strong effect on measurements of galaxy
environment. HIPASS also identifies gas-rich dwarf and low surface
brightness galaxies, often missed by magnitude-limited optical
surveys. In addition, the dark matter halo masses of HIPASS galaxies
have been modelled as a function of \HI-mass \citep{Mo05}, which makes
for a straightforward analysis of the bias and clustering properties
of HIPASS galaxies.

\begin{figure}  
 \vspace{14pc} 
 \includegraphics{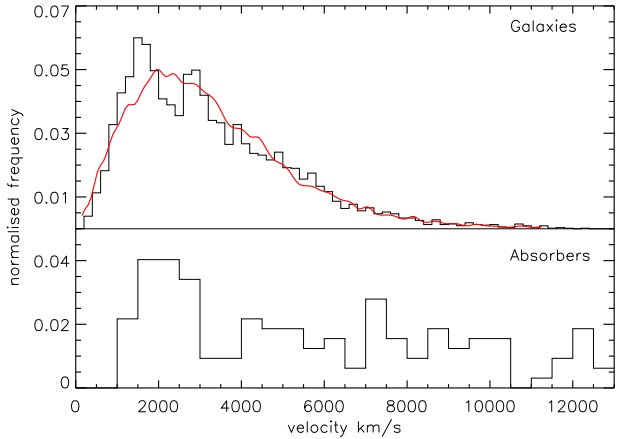} 
\caption{Top panel: Velocity histogram of galaxies in the 
	combined HIPASS catalogues, with the HIPASS selection function 
	overdrawn. Bottom panel:
	Histogram of velocities from the \lya\ absorber literature sample.}
\label{fig:sel_func}
\end{figure}

Random galaxy samples were produced from a Poisson sky distribution
with random velocities weighted by the HIPASS selection function (see
Figure~\ref{fig:sel_func}). The random samples each contain
$2\times10^{4}$ sources.

\subsection{\Lya\ absorbers}
\label{sec:data_lya}
Low redshift, low column density \Lya\ absorbers were selected from
the Space Telescope Imaging Spectrograph \citep[STIS,][]{Bowen02,
Penton04, Williger05} and the Goddard High Resolution Spectrograph
\citep[GHRS,][]{Impey99, Penton00} observations in the
literature. These absorbers were chosen to overlap with the combined
HIPASS catalogues, i.e.  $\delta$$<$+25\degree\ and heliocentric
velocity between 1,000 and 12,700 \kms. The lower limit of 1,000 \kms\
was applied as many lines-of-sight have significant Galactic
absorption features. Only lines with significance level greater than
$4.5\sigma$ were included from \cite{Impey99} and $4\sigma$ from
\cite{Penton00}, consistent with what these authors considered to be
solid detections. The column densities calculated with Doppler
parameter $b=30$ \kms\ are adopted from \cite{Penton00, Penton04}. The
column densities of the absorbers in the literature sample lie in the
range $12.41\leq\log (\nhi/$\cm$) \leq14.81$, although the 10
absorbers from \cite{Impey99} only have equivalent width measurements
(in the range 0.176 to 1.809 \AA). For consistency, each line-of-sight
in the sample contains absorbers from a single reference only (the
most recent reference is used). There are 129 absorbers from 27
lines-of-sight in the full STIS plus GHRS literature sample.


Although a number of low redshift \Lya\ absorbers have been detected
with the lower resolution Faint Object Spectrograph
\citep[FOS,][]{Jannuzi98, Bechtold02}, their equivalent width
distribution and velocity uncertainty is quite different from
absorbers detected with STIS and GHRS. Furthermore, only 11 additional
absorption lines overlap with the HIPASS survey region, thus only
absorbers detected with STIS and GHRS have been included in the
sample.

Random absorber samples were produced by generating random
lines-of-sight on the sky populated with absorbers, where the number
distribution of absorbers per random line-of-sight matched the mean
(4.8) and standard deviation (3.0) of the absorber sample from the
literature. Lines-of-sight with zero absorbers were not included in
either the data or random samples. The random absorber velocities are
weighted by the velocity distribution of the data absorbers. No
evolution is expected or observed in the absorber population over the
HIPASS redshift interval 0$<$z$<$0.04 \citep{Penton04}. The data
absorber sample however does not have a flat velocity distribution (see
Figure~\ref{fig:sel_func}). Ideally the absorber selection function
should be constructed from the sum of redshift path lengths of the 27
lines-of-sight, but this information is not available for all
lines-of-sight. Most (91/129) of the absorbers in the sample are taken
from \cite{Penton00} or \cite{Penton04}. The path length available for
the detection of \Lya\ absorbers for a superset of these sight-lines
\citep[see figure 6 of][]{Penton04} is almost a factor of two higher
at low redshift ($z=0.01$) compared with that at the HIPASS redshift
limit ($z=0.04$). This trend is reflected in the shape of the
histogram of \Lya\ absorber velocities in
Figure~\ref{fig:sel_func}. Since the full absorber sample used in this
paper is a compilation of data from different observing programs and
instruments as well as data reduction, line selection and measurement
techniques, weighting the velocity distribution of the random sample
will minimise systematic effects of the non-uniform velocity
distribution. The final random absorber samples contain around 5,000
absorbers each. Results using the different random absorber samples
show no significant variation within the quoted uncertainties.

\section{Cross Correlation Method}
\label{sec:method}

The cross-correlation function, $\xi(\sigma,\pi)$, is calculated from
the \cite{Davis83} estimator,
\begin{equation}
\xi(\sigma,\pi)=\frac{AG(\sigma,\pi)}{RG(\sigma,\pi)}\frac{n_{RG}}{n_{AG}}-1,
\label{eqn:1}
\end{equation}
where $AG(\sigma,\pi)$ is the number of data absorber--galaxy pairs
with projected separation, $\sigma$, between $\sigma-\delta\sigma/2$
and $\sigma+\delta\sigma/2$, and radial separation, $\pi$, between,
$\pi-\delta\pi/2$ and $\pi+\delta\pi/2$.  $RG(\sigma,\pi)$ is the
number of pairs consisting of a random absorber and a data galaxy. The
projected separation between two objects, $\sigma_{ij}$, is converted
from angular separation, $\theta_{ij}$, by
$\sigma_{ij}=($vel$_i+$vel$_j$)/$H_0\times \tan(\theta_{ij}/2)$. The
separation along the line-of-sight is simply the velocity difference,
$\pi_{ij}=|$vel$_i-$vel$_j|/H_0$. Both $\delta\sigma$ and $\delta\pi$
are set at 0.1 \h\ Mpc.  The sampling of 0.1 \h\ Mpc is used to ensure
a data point exists at reasonably small separations in the correlation
function. Equation~(\ref{eqn:1}) is calculated in the range 0 to 100
\h\ Mpc in the $\sigma$ and $\pi$ directions. The function is
normalised by the number of random pairs, $n_{RG}$, and data pairs,
$n_{AG}$. Only pairs with separations less than 90\degree\ are
included in the calculation (corresponds to 20 \h\ Mpc for the lowest
velocity absorbers).\footnote{Increasing the maximum separation to
180\degree\ produces indistinguishable results for the projected auto-
cross-correlations for $\sigma<10$ \h\ Mpc, variations at larger
$\sigma$ are well within the quoted uncertainties.} The galaxy
auto-correlation function is calculated in a similar way, using the
random galaxy sample. The spherical average of $\xi(\sigma,\pi)$ gives
the redshift-space correlation function, $\xi(s)$, where
$s=\sqrt{\sigma^2+\pi^2}$.


The separations used to calculate the cross-correlation function do
not take peculiar velocities into account. To reconcile the increased
velocity dispersion along the line-of-sight (the finger-of-god effect)
the cross-correlation function is integrated along the velocity axis
to produce a projected correlation function, $\Xi(\sigma)$.
\begin{equation}
\Xi(\sigma)=2\int_0^\infty \xi(\sigma,\pi)d\pi
\label{eqn:2}
\end{equation}
In practice the upper limit of the integral is 50 \h\ Mpc, and other
upper limits are investigated. The projected and real-space
(i.e. $\xi(r)$, distortion free) correlation functions are related via
\begin{equation}
\frac{\Xi(\sigma)}{\sigma}
=\frac{2}{\sigma}\int_\sigma^\infty \frac{r\xi(r)dr}{\sqrt{r^2-\sigma^2}}
\end{equation}
\citep{Davis83}. If a power law functional form is assumed for the
real-space correlation function, $\xi=(r/r_0)^{-\gamma_r}$, then the
parameters $r_0$ and $\gamma_r$ can be derived analytically from the
power-law coefficients of the projected correlation function,
$\Xi(\sigma)/\sigma=A(r_o,\gamma_r)\sigma^{-\gamma_r}$, where 
\begin{equation}
A=r_0^{\gamma_r} \Gamma \left(\frac{1}{2}\right) 
\Gamma \left(\frac{\gamma_r-1}{2}\right) 
/\Gamma \left(\frac{\gamma_r}{2}\right)
\end{equation}
\citep{Davis83}. At large (linear) separations, \cite{Mo96} have used
the PS formalism to show that the real-space auto-correlation function
of galaxies is related to the dark matter correlation function,
$\xi_{DM}$, via the bias, $b$,
\begin{equation}
\label{eqn:xigg}
\xi_{gg}=b^2(M_g)\xi_{DM}.
\end{equation}
This relation has been tested using N-body simulations. An equivalent
relation holds for the cross-correlation function of two populations
of objects \citep[e.g.][]{Mo93}, in this case absorbers and galaxies,
\begin{equation}
\label{eqn:xiag}
\xi_{ag}=b(M_a)b(M_g)\xi_{DM}.
\end{equation}
Thus, the amplitude of cross- to auto- correlation function can be
expressed as the ratio of the bias of absorber dark matter haloes,
$b(M_a)$, to galaxy dark matter haloes, $b(M_g)$.
\begin{equation}
\label{eqn:xiratio}
\frac{\xi_{ag}}{\xi_{gg}}=
\left(\frac{r_{0,ag}}{r_{0,gg}}\right)^{\gamma_{r,gg}}=
\frac{b(M_a)}{b(M_g)}
\end{equation}
As outlined in \cite{Mo02}, the ellipsoidal collapse model of bias
from \cite{Sheth01} is more accurate than the original spherical model,
hence it shall be adopted here. To directly compare the amplitudes of
the cross- and auto-correlation functions, their slopes,
$\gamma_{r,ag}$ and $\gamma_{r,gg}$, must be equal. To ensure this the
coefficients, $r_{0,gg}$ and $\gamma_{r,gg}$, are calculated first,
then $r_{0,ag}$ is determined by fixing
$\gamma_{r,ag}$=$\gamma_{r,gg}$ and using a Levenberg-Marquardt
non-linear least squares fit. For the auto-correlation function the
uncertainties are calculated using jackknife resampling, where the
variance, $\sigma_{\xi}^2$, is measured by calculating the correlation
function N (24) times, with all data and random points from 1 hour of
RA removed each time.
\begin{equation} 
\label{eqn:jk}
\sigma_{\xi}^2=\frac{N-1}{N}
\sum_i \left(\overline{\xi}-\xi_i\right) ^2
\end{equation}
Uncertainties in the cross- and absorber auto- correlation functions
are also calculated using jackknife resampling, with one line-of-sight
removed each time. Jackknife resampling takes into account cosmic
variance within the HIPASS volume and represents the uncertainties more
accurately than $\sqrt{N}$ errors, which are not independent.

In summary, by simply measuring the ratio of the cross- to
auto-correlation functions, together with known halo biases and halo
masses of galaxies, equation (\ref{eqn:xiratio}) can be used to
calculate the mass of haloes in which \Lya\ absorbers are embedded.
This method relies on the assumption that the relative bias can infer
a mass for the absorber halo.

\begin{figure}  
 \vspace{23pc} 
 \includegraphics{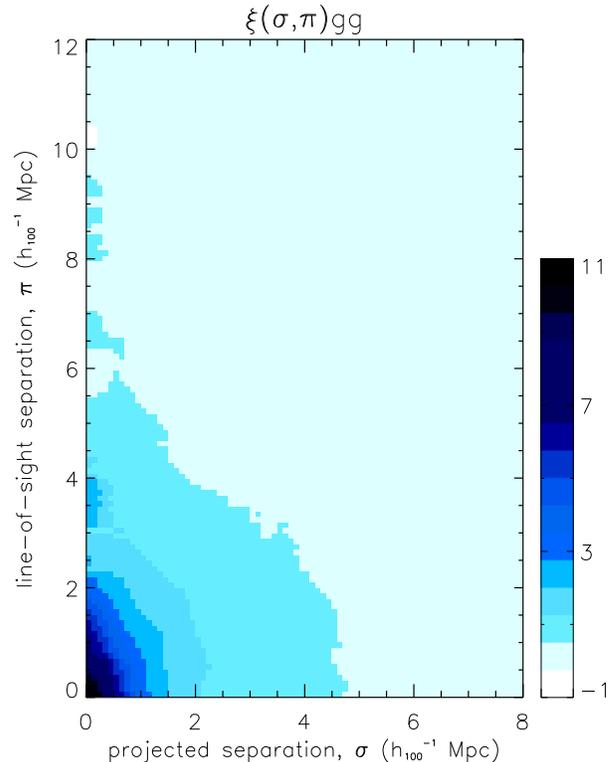} 
\caption{The galaxy auto-correlation function. The function is
calculated with a resolution of $\delta\sigma=\delta\pi=0.1$ \h\ Mpc
and is smoothed with a boxcar width of 9 pixels in this diagram.The
smoothing level of 9 pixels is used to emphasize features on $>$ 1 \h\
Mpc scales. Moderate changes to the sampling and smoothing levels do
not change the these features. Jackknife realisations show typical
pixel variations of the order $\xi_{gg}<0.2$.}
\label{fig:contour_gg}
\end{figure}

\begin{figure}	 
 \vspace{23pc} 
 \includegraphics{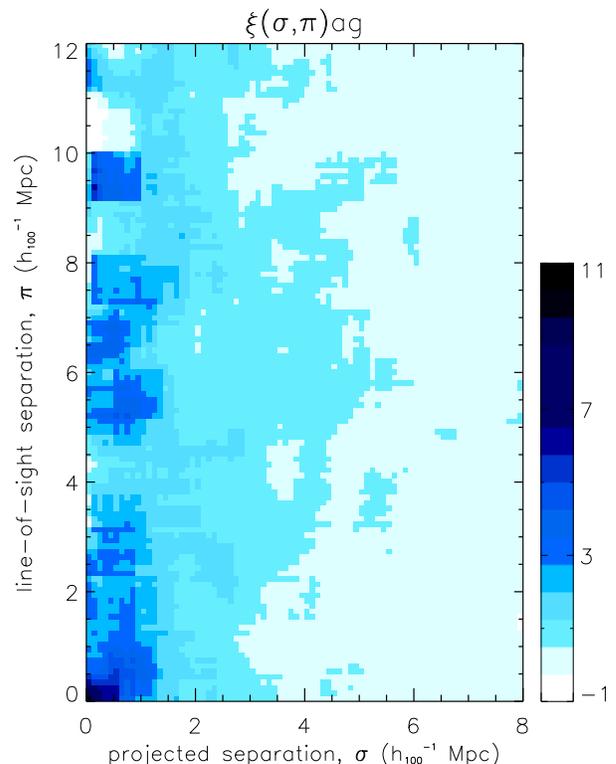} \caption{The absorber-galaxy
 cross-correlation function. Smoothing as per Figure~\ref{fig:contour_gg}.
 Jackknife realisations show typical pixel variations of the order
 $\xi_{ag}<0.5$.}
\label{fig:contour_ag}
\end{figure}

\section{Results}
\label{sec:results}

The auto- and cross- correlation functions, $\xi(\sigma,\pi)_{gg}$ and
$\xi(\sigma,\pi)_{ag}$, are given in Figures~\ref{fig:contour_gg} and
\ref{fig:contour_ag}. The finger-of-god effect due to increased
velocity dispersion along the line-of-sight is apparent for the galaxy
auto-correlation, but is much more exaggerated for the absorber-galaxy
cross-correlation.\footnote{The elongation along the velocity axis is
not dominated by a particular line-of-sight, nor the fact that
absorbers are detected along lines-of-sight, nor the absorbers'
n(z). These possible effects have been investigated by calculating the
cross-correlation along each line-of-sight separately, calculating the
cross-correlation with one absorber from each line-of-sight only,
and considering only absorbers and galaxies with velocities less than
5,000 \kms. In each of these test cases the elongation along the
velocity axis remains.} The same effect is seen in absorber-galaxy
cross-correlation functions from simulations \citep{Dave99}, and is
interpreted as the draining of gas from low-density regions into
collapsed structures. Figures~\ref{fig:contour_gg} and
\ref{fig:contour_ag} also reveal that at very small separations the
auto-correlation signal is stronger, however, the cross-correlation
maintains a higher value over a larger section of $\sigma-\pi$
space. This qualitative comparison suggests that absorbers are not as
highly clustered as galaxies on $<1$ \h\ Mpc scales, though not
avoiding the highest density regions (which are occupied by galaxies),
the normalised chance of finding an absorber at larger scales ($>2$
\h\ Mpc) is more likely than finding another galaxy.

\begin{figure}   
 \vspace{14pc} 
\includegraphics{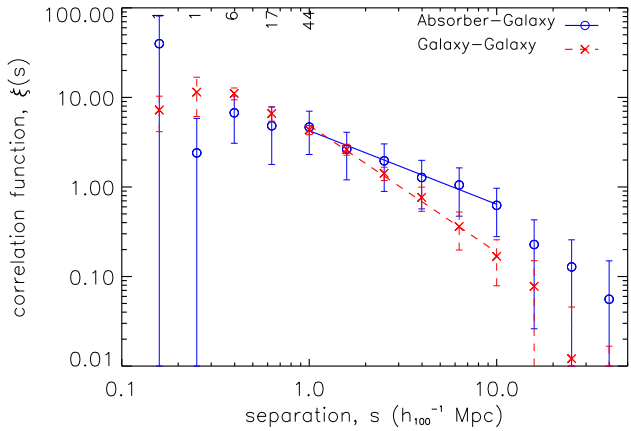} \caption{The absorber-galaxy cross-correlation
 function ($\xi(s)_{ag}$, circles \& solid line) and the galaxy
 auto-correlation function ($\xi(s)_{gg}$, crosses \& dashed line) in
 redshift-space, s. The five numbers indicate the number of absorber-galaxy 
 pairs that contribute to the first five $\xi(s)_{ag}$
 points in the diagram. See text for power-law fit description. Note that
 the y-axis range varies from figure to figure.}
\label{fig:xi_s}
\end{figure}

The redshift-space correlation function, $\xi(s)$, is calculated from
the spherical average of $\xi(\sigma,\pi)$. Both the auto- and
cross-correlation functions are plotted in
Figure~\ref{fig:xi_s}. Power-law fits to the functions
$\xi(s)_{gg}=(s/s_{0,gg})^{-\gamma_{s,gg}}$ and
$\xi(s)_{ag}=(s/s_{0,ag})^{-\gamma_{s,ag}}$ are made over the range
$1-10$ \h\ Mpc for consistency with the projected correlation function
fits (see below). The function parameters are $s_{0,gg}=3.1\pm0.5$ \h\
Mpc, $\gamma_{s,gg}=1.4\pm0.5$, $s_{0,ag}=6\pm4$ \h\ Mpc, and
$\gamma_{s,ag}=0.8\pm0.2$. These results can be compared directly with
\cite{Morris93}. Their pure Hubble flow galaxy auto-correlation
(figure 7a) is marginally higher in amplitude compared with the HIPASS
galaxy auto-correlation function. This is expected due to the weaker
clustering strength of \HI-selected galaxies and is discussed in much
detail in \cite{Meyer03}.\footnote{Meyer uses only the southern HIPASS
catalogue, the \cite{Landy93} estimator and an integration upper limit
of 25 \h\ Mpc, whereas both southern and northern HIPASS catalogues,
the \cite{Davis83} estimator and an integration upper limit of 50 \h\
Mpc are use here. The Meyer results lie within the
uncertainties quoted here.} The absorber-galaxy cross-correlation
function of \cite{Morris93} based on 17 absorbers along the 3C273
line-of-sight is much weaker than the redshift-space correlation
presented here, based on 129 absorbers along 27 lines-of-sight. The
large uncertainty in the cross-correlation amplitude
($s_{0,ag}=6\pm4$) reflects the large variation from one line-of-sight
to another, and emphasises the need for multiple lines-of-sight in
this type of analysis.



\begin{figure}  
 \vspace{14pc} \includegraphics{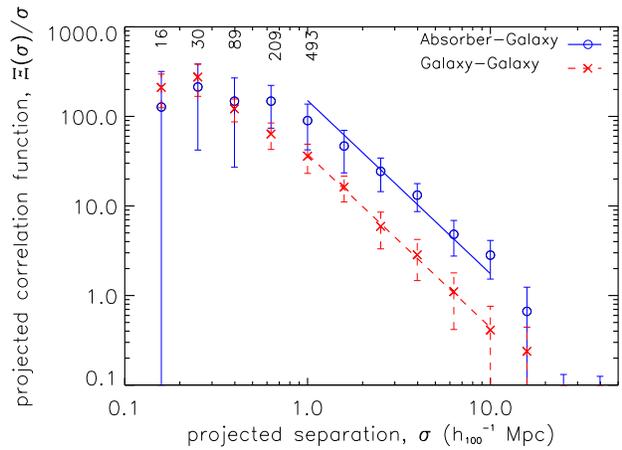} \caption{The projected
 absorber-galaxy cross-correlation function (circles \& solid line)
 and the galaxy auto-correlation function (crosses \& dashed
 line). The five numbers indicate the number of absorber-galaxy pairs
 that contribute to the first five $\Xi(\sigma)_{ag}$ points in the
 diagram.  See text for power-law fit description.}  \label{fig:proj}
\end{figure}

As described in the method, integrating the correlation function along
the velocity axis removes the effect of peculiar velocities along the
line-of-sight and allows parameters of the real-space correlation
function to be calculated. It is clear from
Figure~\ref{fig:contour_ag} that collapsing $\xi(\sigma,\pi)$ along
the $\pi$ axis will produce a stronger signal in the absorber-galaxy
cross-correlation at projected separations greater than 1 \h\ Mpc. The
resulting auto- and cross-correlations are given in
Figure~\ref{fig:proj}, and show that the absorber-galaxy
cross-correlation is indeed stronger than the HIPASS galaxy
auto-correlation for the projected spatial range 1$-$10 \h\ Mpc.

\begin{figure}  
 \vspace{14pc}
 \includegraphics{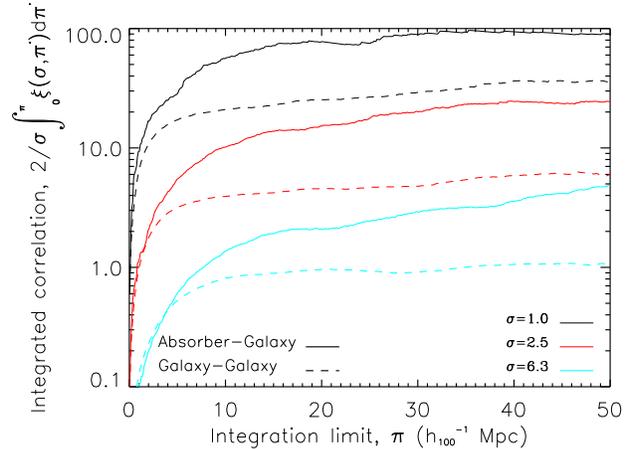}
 \caption{Projected cross- and auto- correlation functions compared at 
   different $\sigma$ values as a function of integration limit, $\pi$, 
   see equation (\ref{eqn:2}).}
 \label{fig:pi_limit_compare}
\end{figure}

To test the robustness of the projected correlation function, the
upper integration limit in Equation (\ref{eqn:2}), $\pi$, is varied
between 0 and 50 \h\ Mpc. Both the auto- and cross- correlation
functions are relatively insensitive to the choice of integration
limit. Lowering the upper limit of the integral does lower the
amplitude of both correlation functions slightly, however
$\Xi(\sigma)_{ag}$ remains greater than $\Xi(\sigma)_{gg}$ over a
large area of parameter space. This trend is apparent in
Figure~\ref{fig:pi_limit_compare} where the solid line
(cross-correlation) remains above the dashed line (auto-correlation)
for displayed values of $\sigma$ (different shades).  Galaxy--galaxy
pairs with radial separations greater than 50 \h\ Mpc account for only
6 per cent of the total number of data pairs in the auto-correlation
calculation.

\begin{figure}  
 \vspace{14pc}
 \includegraphics{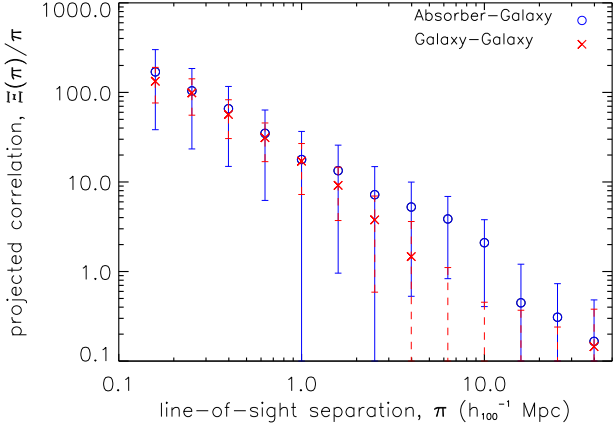}
 \caption{The line-of-sight absorber-galaxy cross-correlation function
	(circles) and the galaxy auto-correlation function (crosses).}
 \label{fig:projpi}
\end{figure}

Parameters of the projected and real-space correlation functions are
calculated by the method described in section~\ref{sec:method}. The
power-law fit is only made over the range $1-10$ \h\ Mpc since the
cross-correlation flattens significantly at small separations and the
uncertainties are high at large separations. A correlation length of
$r_{0,gg}=3.5\pm0.7$ \h\ Mpc and slope of $\gamma_r=1.9\pm0.3$ is
found for the auto-correlation, and $r_{0,ag}=7.2\pm1.4$ \h\ Mpc for
the cross-correlation. These parameters are plotted on the figure as
$\Xi(\sigma)/\sigma=A(r_o,\gamma_r)\sigma^{-\gamma_r}$ for both the
auto- and cross- correlation functions, fixing
$\gamma_{r,ag}=\gamma_r$ for reasons described in the method. Using
equation (\ref{eqn:xiratio}) these parameters yield a bias ratio of
$4.0\pm1.3$. The geometric mean \HI\ mass of galaxies contributing to
pairs in the range $1\leq\sigma\leq10$ \h\ Mpc is log(\mhi/\msun) =
8.8 $h^{-2}_{100}$, with a geometric standard deviation of 1.1. This
corresponds to a halo mass of log($M$/\msun) = 11.0 \h\ with
1-$\sigma$ lower and upper limits of log($M$/\msun) = 10.8 and 11.6
respectively \citep{Mo05}. Combining this mean galaxy halo mass,
together with the bias ratio calculated above and the linear bias
relation \citep{Mo02} gives a value of log($M$/\msun) $=14.2$ \h\
(with 1-$\sigma$ lower and upper limits of 13.6 and 14.5 respectively)
for the mass of haloes in which \Lya\ absorbers are embedded. It
should be noted that log(\mhi/\msun) = 8.8 $h^{-2}_{100}$ is lower
than the geometric mean \mhi\ of all HIPASS galaxies because the
cross-correlation function is dominated by galaxies located nearby,
which on average are lower in \HI\ mass. Nevertheless the bias
relation at $z=0$ is a relatively flat function of halo mass for halo
masses less than $10^{12}$ \msun, so the choice of $M_g$ has little
effect on the final result.

\begin{figure}  
 \vspace{14pc}
 \includegraphics{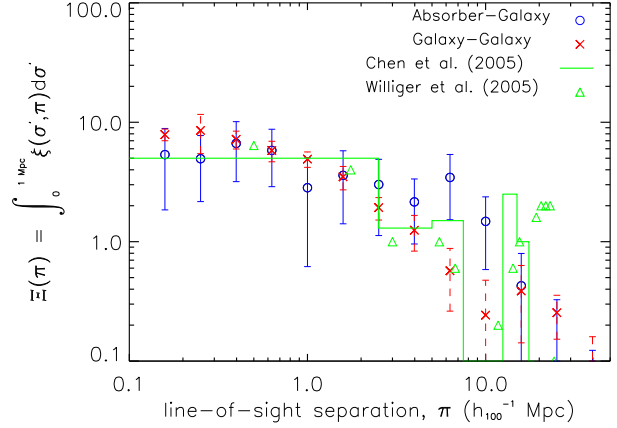}
 \caption{The line-of-sight absorber-galaxy cross-correlation function
	(circles) and the galaxy auto-correlation function (crosses) 
	integrated to $\sigma=1$ \h\ Mpc only to compare with 
	the cylindrical-style cross-correlation functions from 
	Chen et al. (2005) and Williger et al. (2005).}
\label{fig:projpi_1Mpc}
\end{figure}

The integral of $\xi(\sigma,\pi)$ along the $\sigma$ axis is included
for comparison with single line-of-sight absorber-galaxy
cross-correlation functions that are expressed in terms of velocity
separation. An integration limit of $\sigma=50$ \h\ Mpc is used. This
function is given in Figure~\ref{fig:projpi}, the large error bars
reflect the significant variations from one line-of-sight to another
in the already weakened correlation signal along the velocity axis due
to the finger-of-god effect. As noted in \cite{Williger05} this
calculation produces auto- and cross- correlation functions of similar
amplitudes. Direct comparisons can be made with the cylindrical-style
$\xi(\Delta v)$ functions of \cite{Chen05} and \cite{Williger05} by
integrating to $\sigma=1$ \h\ Mpc (although Williger et al. have a
projected separation limit that increases from 1 to 1.6 \h\ Mpc along
the redshift cylinder) producing the function,
\begin{equation}
\Xi(\pi)=\int_0^{\sigma=1 Mpc} \xi(\sigma^\prime,\pi)d\sigma^\prime,
\end{equation}
given in Figure~\ref{fig:projpi_1Mpc}. Interestingly the
cross-correlation drops below the auto-correlation function for
$\sigma<1$ \h\ Mpc in agreement with \cite{Chen05} and
\cite{Williger05}, however within the error bars, most points overlap.

\begin{figure}  
 \vspace{14pc}
 \includegraphics{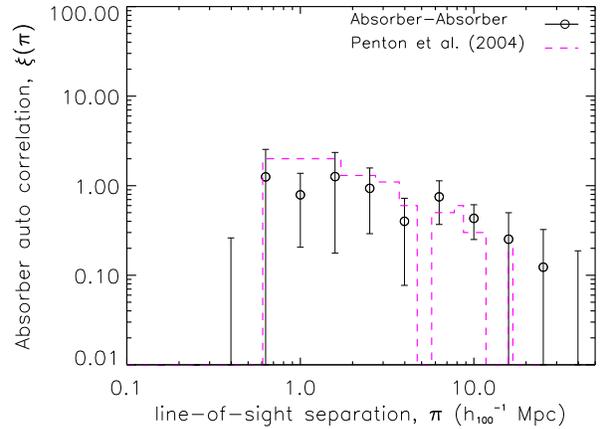}
 \caption{Absorber auto-correlation function (circles) with 
	Penton et al. (2004) two-point correlation function, 
	$\xi(\Delta v)$, for comparison.}
 \label{fig:xi_aa}
\end{figure}

Although the mean number of absorbers per line-of-sight is only 4.8, it
is still possible to measure the absorber auto-correlation in
individual lines-of-sight,
\begin{equation}
\xi(\pi)=\frac{AA(\pi)}{AR(\pi)}\frac{n_{AR}}{n_{AA}}-1,
\end{equation}
where $AA(\pi)$ is the number of data absorber pairs with separation,
$\pi$, and $AR(\pi)$ is the number of random absorber pairs. Random
absorbers are generated as described in Section~\ref{sec:data_lya}.
Lines-of-sight with a single absorber are not included in the
calculation.  Figure~\ref{fig:xi_aa} shows that absorbers correlate
with themselves very weakly, in agreement with \cite{Penton04}, shown,
and also \cite{Impey99} and \cite{Williger05}. This agreement is
anticipated as these absorbers are drawn from those data sets, but
still forms a good consistency check as these absorbers represent the
first few in each line-of-sight, so there is less opportunity to form
pairs.

\begin{figure}  
 \vspace{14pc} 
\includegraphics{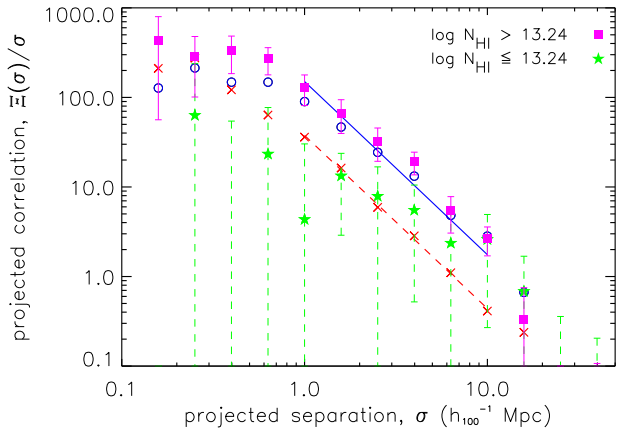} \caption{The projected absorber-galaxy
 cross-correlation function for absorbers with log (\nhi/\cm) $>13.24$
 and log (\nhi/cm) $\leq13.24$.  The cross-correlation from the full absorber 
 data set and galaxy auto-correlation functions from Figure~\ref{fig:proj} 
 are also plotted for comparison (circles with solid line and 
 crosses with dashed line respectively).}
 \label{fig:proj_nhi}
\end{figure}

To test the relationship between absorber column density and
correlation strength with galaxies, the absorber sample is split into
two subsets defined by the median column density, log (\nhi/\cm)
$=13.24$. For consistency with \cite{Penton02,Penton04} a Doppler
parameter of 30 \kms\ is adopted for absorption lines from
\cite{Impey99} to convert the quoted equivalent width measurements
into column densities. Random absorber catalogues were re-generated
for these sub-samples based on the new velocity distribution, and new
mean and standard deviation of each
line-of-sight. Figure~\ref{fig:proj_nhi} gives the cross-correlation
for absorbers above and below the median \nhi\ and shows definite
evidence for segregation of lower and higher column density
systems. The cross-correlation for absorbers with log (\nhi/\cm)
$>13.24$ is somewhat stronger than that from the full sample of
absorber at projected separations less than 1 \h\ Mpc, although the
uncertainties overlap. Low column density absorbers (log (\nhi/\cm)
$\leq13.24$) show a weaker correlation with galaxies, however the
uncertainties are large and the results are consistent with no
correlation on scales less than 1 \h\ Mpc.

\section{Discussion}
\label{sec:dis}

The locations of low-\nhi\ \Lya\ absorbers and gas-rich galaxies have
been cross-correlated at $z=0$, giving a redshift-space clustering
signal similar in strength to that of galaxy self-clustering.  The
projected cross-correlation removes the increased velocity dispersion
along the line-of-sight and from it the real-space clustering can be
derived. The cross-correlation function is a powerful statistic as it
can be related to the underlying dark matter distribution. Here the
absorber cross-correlation results are discussed in relation to
mini-haloes, galaxy groups, large-scale filaments, numerical
simulations and the environs of gas-rich galaxies.

\subsection{Ruling out mini-haloes}

A typical mini-halo has a circular velocity, $v_c$, of 30 \kms\
\citep[e.g.][]{Mo94}, which corresponds to a halo mass and bias of
$6.8\times10^9$ \msun\ and 0.73 respectively. Using
Equation~\ref{eqn:xiratio} and the auto-correlation result, such
haloes would have a cross-correlation length of 3.3 \h\ Mpc. Comparing
this correlation length with the absorber cross-correlation length
($r_{0,ag}=7.2\pm1.4$ \h\ Mpc) rules out mini-haloes for the
confinement of low-\nhi\ \lya\ absorbers at $z=0$ at the 99 percent
confidence limit.

\subsection{Comparison with galaxy groups}

The ratio of real-space clustering strengths inferred from the
projected correlation functions imply that \Lya\ absorbers are
embedded in haloes with masses in the range $13.6<$ log($M$/\msun)
$<14.5$ \h. This halo mass range is of the same order as the median
dynamical mass of galaxy groups in HIPASS, log($M$/\msun) $\sim13.8$
\h\ \citep{Stevens05}. This similarity warrants a careful comparison
between correlation properties of absorbers and groups.

The cross-correlation between groups and galaxies has been measured
with both the Sloan and 2dF galaxy redshift surveys \citep{Yang05}.
The most massive groups (halo mass log($M$/\msun) $\le13.8$ \h)
display the most elongation along the $\pi$ axis due to larger
velocity dispersions, similar to that seen in the absorber
cross-correlation. The shape of the group-galaxy cross-correlation
resembles that of the absorber-galaxy cross-correlation at $\sigma<2$
and $\pi<10$ \citep[][ figure 1 -- all groups and faintest
galaxies]{Yang05}, although the group-galaxy correlation level is
higher. No 2-D flattening signature of peculiar motions due to infall
is seen in the absorber-galaxy cross-correlation, whereas this
signature is apparent for groups. The projected correlation is similar
in slope at $1<\sigma<10$, but the flattening of the
absorber-galaxy correlation, leads to a much stronger group-galaxy
correlation at $\sigma<1$.

The main difference between the correlation properties of absorbers
and groups is their self-clustering. Absorber self-clustering (see
Figure~\ref{fig:xi_aa}) appears to be weak. The weak clustering could
be due to the measurement technique, since absorber self-clustering
measurements are restricted to the line-of-sight axis and the
clustering signal is elongated, and therefore diluted along this
axis. The cross-correlation measured along the redshift axis
has almost three times the amplitude as that measured perpendicular to
the line-of-sight (compare Figures~\ref{fig:proj} and
\ref{fig:projpi}). Interestingly if the absorber auto-correlation
(Figure~\ref{fig:xi_aa}) were increased by a factor of three it would
be comparable to the redshift-space galaxy auto-correlation
(Figure~\ref{fig:xi_s}). N-body simulations and analytic models can be
used to measure absorber self-clustering in real-space. At least in
the case of the warm-hot intergalactic medium (WHIM, slightly higher
column densities than the absorbers considered in this paper) the
real-space clustering of absorbers at $r=1$ \h\ Mpc could be a factor
two less \citep{Valageas02} or two more \citep{Dave01} than that of
galaxies. 

Groups of galaxies on the other hand are strongly self-clustered, but
not necessarily as strong as their constituent galaxies. More massive
haloes are more strongly clustered and contain more galaxies per
halo. Thus there are many more galaxies than haloes in strongly
clustered environments, and a comparable number of galaxies and haloes
in the field. Therefore, purely by relative numbers, galaxies in
strongly clustered environments will dominate their correlation
function, whereas the correlation signal of groups will be more evenly
weighted. This weighting can lead to the self-clustering signal of
groups being {\it{weaker}} than that of their constituent galaxies.
Naturally, group self-clustering depends on the mass of haloes
considered. Padilla et al. (\citeyear{Padilla04}) show that the
redshift-space correlation length of groups identified in the 2dF
galaxy redshift survey increases in amplitude from $s_0=5.5$ \h\ Mpc
for the entire sample, to 12.6 \h\ Mpc for groups with a median mass
of $1.1\times10^{14}$ \h\ \msun (compared with $s_0=6.82$ \h\ Mpc for
2dF galaxies from Hawkins et al. \citeyear{Hawkins03}). The projected
auto-correlation of this larger mass group sample is quite similar in
amplitude and shape to the \cite{Yang05} group cross-correlation
function described above. Hence it appears that the self-clustering
properties of groups are likely to be strong, but could be weaker than
galaxies. On the other hand, the self-clustering properties of \lya\
absorbers are likely to be weak, but could be as strong as galaxies.

Absorbers, galaxies and groups each trace the underlying dark matter
halo distribution in different ways. By comparing the clustering
properties of absorbers to that of other objects, the dark matter
haloes in which absorber reside can be revealed. As found by many
authors (see Section~\ref{sec:intro}), the possibility that \lya\
absorbers arise in individual galaxy haloes is unlikely. At a distance
of 20 \h\ Mpc, HIPASS is sensitive to galaxies with \mhi\ $\geq
2\times10^{8}$ $h^{-2}_{100}$ \msun. Considering only galaxies and
absorbers within this distance, the nearest galaxy to 50 per cent of
absorbers is greater than 1 \h\ Mpc in redshift-space, clearly not
within the halo of an individual galaxy. Galaxy groups give a
$\xi(\sigma,\pi)$ diagram that is similar in shape to the
absorber-galaxy cross-correlation. In addition, the ratio of
real-space clustering lengths imply that \lya\ absorbers are embedded
in haloes with masses similar to that of galaxy groups. These
similarities provide some evidence for the notion that \Lya\
absorbers are associated with galaxy group environments.


\subsection{Nature of the correlation signal compared with simulations}

Taking the projected absorber-galaxy cross-correlation
(Figure~\ref{fig:proj}) at face value, it can be concluded that
stepping 1 to 10 \h\ Mpc away from an \HI-selected galaxy, the
normalised chance of finding an absorber is more likely than finding
another galaxy. This likelihood is in addition to the fact that \lya\
absorbers are $\sim$40 times more frequent than galaxies at zero
redshift.\footnote{The number density of absorbers with $13.1<$ log
(\nhi/\cm) $<14.0$ is dN/dz=1.85 \citep{Penton04}, the value for
gas-rich galaxies is dN/dz=0.045 \citep{Zwaan05}.} The exact covering
factor of neutral gas 1 \h\ Mpc away from a galaxy however requires a
value for the size of the absorbers, or an assessment of
non-detections. At smaller separations the cross-correlation flattens
with respect to the galaxy auto-correlation, indicating that at $<1$
\h\ Mpc the normalised chance of encountering a low-\nhi\ \lya\
absorber is less than that of another galaxy - perhaps giving way to
higher column density systems. Flattening of the absorber
auto-correlation function at a few hundred kiloparsecs is also seen in
analytic models and N-body simulations of WHIM absorbers
\citep{Dave01, Valageas02}. This length ($\sim$ 300 kpc) corresponds
to the characteristic thickness of the WHIM filaments. A fairer
comparison can be made with $\xi(\sigma,\pi)$ diagrams for the
cross-correlation of galaxies with low column density absorbers in
\cite{Dave99}. Their results suggest the correlation function turns
over at small separations, but the interpretation is limited by the
resolution of the simulations. The results presented here show that
the cross-correlation flattens at separations less than 600 \h\
kpc. It is possible that this separation corresponds to a
characteristic thickness of the large-scale structures in which low
column density \lya\ absorbers are embedded.


Modelling the low redshift Universe is a complicated business.
Ideally the results presented here should be compared with directly
with $\xi(\sigma,\pi)$ and $\xi(r)$ from simulations and models at
zero redshift. Future simulations may be able to distinguish the
signature of absorbers embedded in different large-scale structures,
for example that of groups or filaments. The essential effect that
needs to be tested is whether the elongation along the velocity axis
seen in the cross-correlation is due to absorbers that are embedded
in a large gravitational potential. Another important insight offered
by simulations and models is measuring absorber self-clustering in
real-space. Given the large elongation of the absorber-galaxy
cross-correlation in redshift-space, and the limitation that absorber
self-clustering currently can only be measured along the
line-of-sight, it is possible that the self-clustering strength of
\lya\ absorbers has been understated.

\subsection{Connection with gas-rich galaxies}

Although the HIPASS galaxies are simply used as test particles to
represent the underlying density of matter, since they are gas-rich
galaxies, gas-rich environments are deliberately
selected. \cite{Stevens04} have shown that the \HI\ content of
galaxies in groups detected in HIPASS is no different to field
galaxies, in contrast with optically selected groups, which are
\HI-deficient \citep{Verdes01}. Thus the intragroup medium of
\HI-selected groups is expected to have a lower ionization fraction,
leading to a higher number of \lya\ absorbers. This may explain why
\cite{Impey99} find that absorbers in the vicinity of the Virgo
cluster lie preferentially in regions of intermediate (rather than
high) galaxy density, since the outskirts of galaxy clusters are known
to be in excess of \HI-rich galaxies by a factor of three (Waugh et
al. \citeyear{Waugh02}).  \cite{Stocke05} find that only 60 per cent
of \lya\ absorbers with $\log(\nhi/$\cm$)\gtrsim 13.5$ (corresponding
to detected \ovi\ absorption lines) arise in optically selected galaxy
groups. Although the difference between optical and \HI-selected
groups in this case may be due to different sensitivity limits as
\cite{Stocke05} consider galaxy groups with more than one L* galaxy,
whereas HIPASS is sensitive to much fainter galaxies (Doyle et
al. \citeyear{Doyle05}).

All galaxies in a representative sample of HIPASS galaxies (Meurer et
al. \citeyear{Meurer05}) have been detected in \Ha\ emission, and are
therefore undergoing star formation. These results emphasize that
interstellar hydrogen is a co-requisite for star formation and that
HIPASS galaxies can be categorised as emission line
galaxies. Interestingly, \cite{Chen05} finds a stronger \lya\
cross-correlation for emission line galaxies, compared with absorption
line galaxies, a result that suggests the environs of gas-rich
galaxies are preferred by \Lya\ absorbers.

Despite the large number of absorbers and lines-of-sight, this study
is limited by the shallow redshift coverage of HIPASS. Jackknife
errors take into account cosmic variance within the HIPASS volume, but
not outside the volume. Infrared galaxy counts show that the local
area ($\sim$200 \h\ Mpc in linear extent, to a distance of $\sim$150
\h\ Mpc) around the south Galactic pole is underdense by 30 per cent
\citep{Frith03}.  \cite{Lin00} suggest a minimum redshift path length,
$\Delta z$, of 10 for statistics that discriminate between absorbers
arising in galactic haloes and those arising in the comic web. The
literature sample used in this paper has $\Delta z\sim1$ only.
HIPASS galaxies also have different clustering properties to optically
selected galaxies \citep{Meyer03}

For all these reasons it would certainly be worth repeating this study
with other galaxy surveys.  Future observations designed to measure
absorber-galaxy clustering to the same extent as seen here would
require a survey of galaxies to projected distances of 10 \h\ Mpc,
i.e. for nearby absorbers with velocities around 10,000 \kms, a radius
of almost 6 degrees would be required around each quasar, with a depth
of at least 100 \h\ Mpc along the line-of-sight.

\section{Summary}
\label{sec:summary}

In this paper 129 low column density (\nhi\ $<10^{15}$ \cm) \lya\
absorbers along 27 lines-of-sight have been cross-correlated with
5,317 gas-rich galaxies at $z=0$. A positive association between
absorbers and galaxies is found to separations of at least 10 \h\ Mpc
in redshift space. The redshift-space auto- and cross-correlation
functions are found to be similar in strength, with correlation
lengths $s_{0,gg}=3.1\pm0.5$ and $s_{0,ag}=6\pm4$ \h\ Mpc
respectively. These results are contrary to that of \cite{Morris93}
based on 17 absorbers along a single line-of-sight, which found a
weaker cross-correlation (the uncertainty on $s_{0,ag}$ is large
however, so a weaker correlation for absorbers can only be ruled out
at the 53 percent confidence limit). The cross-correlation is found to
depend on the column density of the absorbers: higher column density
systems (log (\nhi/\cm) $>13.24$) are more strongly correlated with
galaxies, especially at projected separations less than 1 \h\ Mpc.

On a one-to-one basis, out to a distance of 20 \h\ Mpc, 50 per cent of
\lya\ absorbers are located within 1 \h\ Mpc of a galaxy, and the
other half are located $1-3$ \h\ Mpc from the nearest galaxy. These
results agree with many other studies \citep{Bowen96, Bowen02,
Tripp98, Impey99, Penton02, Cote05} that the majority of \lya\
absorbers do not arise in the haloes of individual galaxies. Bearing
this result in mind, the cross-correlation function can then be used
to measure the extent to which absorbers and galaxies are related.

The amplitude of the absorber-galaxy cross-correlation depends on how
it is integrated. The absorber-galaxy $\xi(\sigma,\pi)$ diagram is
very elongated along the line-of-sight. Integrating along the
projected separation ($\sigma$) axis produces auto- and
cross-correlations with comparable amplitudes. Restricting this
integral to 1 \h\ Mpc leads to a cross-correlation that is slightly
weaker than the auto-correlation at small separations. Thus results
from cylindrical-style cross-correlation functions \citep{Chen05,
Williger05} must be interpreted with caution.

A more conventional approach (as employed in galaxy clustering
analysis) is to integrate along the line-of-sight ($\pi$) axis to
remove the increased velocity dispersion and from it derive the
real-space clustering, and underlying dark matter distribution. The
absorber-galaxy cross-correlation has been measured in real-space for
the first time here. A correlation length of $r_{0,ag}=7.2\pm1.4$ \h\
Mpc is found, compared with $r_{0,gg}=3.5\pm0.7$ \h\ Mpc and slope of
$\gamma_r=1.9\pm0.3$ for the galaxy auto-correlation. Mini-halo models
on the other hand predict a weaker cross-correlation. The
interpretation of these results relies on the assumption that the
increased velocity dispersion seen in the cross-correlation is due to
a larger gravitational potential. These results rule out mini-halo
models for the confinement of low-\nhi\ \lya\ absorbers at $z=0$ at
the 99 percent confidence limit. The ratio of real-space clustering
strengths, together with the bias relation (Equation~\ref{eqn:xiratio})
imply that \Lya\ absorbers are embedded in haloes with masses in the
range $13.6<$ log($M$/\msun) $<14.5$ \h, similar to that of galaxy
groups. The cross-correlation of massive groups with galaxies produces
a $\xi(\sigma,\pi)$ diagram similar in shape to that of the
absorber-galaxy cross-correlation. Thus on average \lya\ absorbers
tend to follow the same cosmic web of material that embeds both
galaxies and groups. This picture is supported by careful analysis of
individual systems \citep{Bowen02}. It is also possible that absorbers
arise in filaments of large-scale structure, although the
cross-correlation amplitude of filaments has not been measured using
observations nor predicted from models or simulations.

The flattening of the cross-correlation function at projected
separation less than $\sim$ 600 \h\ kpc could correspond to a
characteristic thickness of the filaments in which \lya\ absorber are
embedded. A large discrepancy exists between individual \lya\ absorber
sizes derived from photo-ionization models using metal lines
associated with each absorber \citep[$10-70$ pc,][]{Rigby02, Tripp02}
and their minimum transverse sizes derived from quasar line-of-sight
pair analysis \citep[of order 300 \h\ kpc,][]{Penton02,
Rosenberg03}. In agreement with \cite{Rosenberg03}, this discrepancy
could be reconciled if the minimum transverse size corresponds to the
thickness of the filament rather than the individual absorber
size. Furthermore, \cite{Penton02} identify a filament consisting of
eight absorbers and five galaxies at least 20 $h_{70}^{-1}$ Mpc across
and 1 $h_{70}^{-1}$ Mpc thick. This example fits with the $\sim$ 600
\h\ kpc filament thickness inferred from the cross-correlation
function.

The cross-correlation results presented in this paper provide indirect
evidence that \lya\ absorbers arise preferentially in gas-rich galaxy
groups and filaments. Over half the galaxies in HIPASS occur in groups
\citep{Stevens05} and gas with $\log(\nhi/$\cm$)>13.2$ covers 50 per cent
of galaxy filaments \citep{Stocke05} -- a picture consistent with the 
ubiquitous nature of \lya\ absorbers.

\section*{Acknowledgments}

I wish to acknowledge Lister Staveley-Smith \& Rachel Webster who
encouraged an early version of this work, Martin Zwaan for providing
the HIPASS selection function and Nicolas Bouch\'e for providing $z=0$
bias values. This paper would not have been possible without the
efforts of the HIPASS team over the last 10 years from survey
conception to catalogue completion.  I would also like to thank Paul
Hewett and Matteo Viel for useful discussions and Nicolas Bouch\'e,
Michael Murphy and Max Pettini for comments on the manuscript. The
referee, Simon Morris, is also acknowledged for helpful comments that
improved the paper. The work in this paper was supported by a PPARC
rolling grant at the University of Cambridge.

\bibliographystyle{mn2e}
\bibliography{mn-jour,corr_hipass}

\bsp

\label{lastpage}

\end{document}